\documentclass[onecolumn,showpacs,twoside,superscriptaddress,pra]{revtex4}
\usepackage{amssymb, amsbsy, amsmath, latexsym, dsfont, array, layout, graphicx,mathrsfs,color}

\newcommand{\one}{\mathds{1}}
\newcommand{\ket}[1]{\left|{#1}\right\rangle}
\newcommand{\bra}[1]{\left\langle{#1}\right|}

\begin{document}

\title{Two quantum walkers sharing coins}
\author{Peng Xue}
\affiliation{Department of Physics, Southeast University, Nanjing
211189, China}
\author{Barry C. Sanders}
\affiliation{Institute for Quantum Information Science, University
of Calgary, Alberta T2N 1N4, Canada}
\date{\today}

\begin{abstract}
We consider two independent quantum walks on separate lines
augmented by partial or full swapping of coins after each step. For
classical random walks, swapping or not swapping coins makes little
difference to the random walk characteristics, but we show that
quantum walks with partial swapping of coins have complicated yet
elegant inter-walker correlations. Specifically we study the joint
position distribution of the reduced two-walker state after tracing
out the coins and analyze total, classical and quantum correlations
in terms of mutual information, quantum mutual information,
and measurement-induced disturbance. Our analysis shows
intriguing quantum features without classical analogues.
\end{abstract}


\maketitle

\section{Introduction}

Quantum walks (QWs)~\cite{ADZ93} are especially interesting because
of their importance in quantum algorithms
research~\cite{AAKV01,Ken06} and also because they represent an
intriguing quantum version of the ubiquitous classical phenomenon of
random walks (RWs). Originally conceived as a single walker whose
left or right step on a line is entangled with the outcome of
flipping a single two-sided quantum coin, the QW has been extended
to more general cases of higher-dimensional
walks~\cite{MBSS02,FGB11,FGMB11}, multiple walkers and/or multiple
coins~\cite{BCA03,BCAprl,SBKKJ11,RSSJS11,AVWW11,AAM+11,BW11}. These
generalizations enable exploration of QWs in complex settings, which
could connect with real-world phenomena such as transport in
photosynthetic protein complexes~\cite{MRLG08}. Additionally the
inclusion of more walkers and/or coins can efficiently enlarge the
Hilbert space being explored because a linear increase in degrees of
freedom exponentially enlarges the Hilbert space.

Here we explore the complex case of two walkers each carrying and
(quantumly) flipping separate coins but with the freedom to swap
partially the coins between flips. By partial swapping we mean that
the two walkers can effect the unitary operation SWAP$^\tau$ on
their coins: for $\tau=0$, this corresponds to no swapping
whatsoever so the two walkers evolve completely independently, and
the other extreme is $\tau=1$ corresponding to completely swapping
the coins after every step. For~$\tau>0$ a consequence of full or
partial swapping is to cause entanglement between the walkers'
position and coin degrees of freedom. Classically swapping coins
does not change the walker dynamics because each coin flip is
random, but quantumly the effect of swapping is important because of
the unitary dynamics.

Partial or full swapping of coins is interesting as this simple
procedure turns independent QWs into entangled QWs. As multi-walker
QWs could occur in nature, our work suggests new phenomena to
explore in QW behavior. Although we focus on the two-walker case,
the multiple-walker case can be obtained by a natural extension of
our foundational work on this area. Full and partial swapping of
coins between independent QWers is also appealing practically as an
implementation. Each independent walker's can correspond to the
phase of a cavity field, and each walker's two-sided coin is an atom
in the cavity with a superposition of ground and excited electronic
energy states coupled to the cavity via a dispersive nonlinear phase
shift~\cite{TM02,XS08,XSBL08}. The atoms can interact via a
dipole-dipole interaction~\cite{HM09,DRKB02} between (or during)
steps, and the strength of this dipole-dipole interaction determines
the value of~$\tau$.

Two-walker QWs are compared to two-walker RWs by examining the joint
position distributions of the two walkers after tracing out the coin
states. Classically, for any value of~$\tau$, no correlation is
expected and the diffusion of position is marked by its spread~$\sigma$ (standard deviation
of the position distribution) increasing as~$\sqrt{t}$ for~$t$
the elapsed time (which can be expressed as a continuous or a
discrete quantity and is the discrete number of steps in our study).
Quantum walks can be distinguished by ballistic spreading, where the $\sigma$ grows linearly with $t$~\cite{Kem03}
or by Anderson-type localization where the walker's spread becomes constant at large~$t$~\cite{KLMW07,SCP+11}.
This localization effect is due to the walker effectively becoming trapped due to irregularities of the spacing of the lattice traversed by the walker.
Thus, there are three domains of spreading in the asymptotic large~$t$ limit:
ballistic ($\sigma\propto t$), diffusive ($\sigma\propto\sqrt{t}$) and localized~($\sigma\sim$constant).

Here we are interested not only in how the walker's position spreads but also in the correlations between two walkers' positions after tracing out the coins.
To study the correlations, we go beyond evaluating joint position distributions
for different values of~$t$ to studying two-time correlation
functions. Such correlation functions prove to be quite useful for
studying two-walker dynamics with partial coin swapping.

\section{Methods}
\subsection{One walker}

For a single-walker QW on a line, the Hilbert space is
\begin{equation}
\label{eq:H}
    \mathscr{H}=\mathscr{H}_\text{w}\otimes\mathscr{H}_\text{c}
\end{equation}
with the walker Hilbert space~$\mathscr{H}_\text{w}$ spanned by the
orthogonal lattice position vectors~$\{\ket{x}\}$ and
$\mathscr{H}_\text{c}$ the two-dimensional coin space spanned by two
orthogonal vectors which we denote~$\ket{\pm 1}$. Each step by the
walker is effected by two subsequent unitary operators: the
coin-flip operator
\begin{equation}
\label{eq:C}
    C=H=\frac{1}{\sqrt{2}}\begin{pmatrix}1&1\\1 & -1\end{pmatrix},
\end{equation}
for~$H$ the Hadamard matrix and the conditional-translation operator
\begin{align}
\label{eq:S}
    S=&\sum_x\ket{x+1}\bra{x}\otimes\ket{1}\bra{1}
        \nonumber\\&+\sum_x\ket{x-1}\bra{x}\otimes\ket{-1}\bra{-1}.
\end{align}
The resultant step operator is $U=S(\one\otimes C)$ with~$\one$ the
identity operator on~$\mathscr{H}_\text{w}$.

The choice of initial state~$\big|\psi(t=0)\big\rangle$ is important
in studies of QWs because the interference features sensitively
depend on the choice of state. This sensitivity is persistent
because the dynamics are unitary hence do not die out. On the other
hand the general properties of QWs do not depend on the choice of
initial state so the choice of initial state is not crucial provided
that the focus is on such characterization.

As we are interested in general properties, the initial state is not
important so we choose the initial product state with the walker at
the origin of the two-dimensional lattice and hold separate coins in
an equal superposition of the~$+1$ and~$-1$ states:
\begin{equation}
   \big|\psi(t=0)\big\rangle=\frac{1}{\sqrt{2}}\ket{0}(\ket{1}+i\ket{-1}).
\label{eq:ini}
\end{equation}

The differences between QWs and RWs are due to interference effects
(or entanglement) between coin and walkers after several
applications of $U$ (with the number of applications~$t$ being the
discrete time variable). We perform a measurement at some point in
order to know the outcome of the walk. The measurement of the
walker's position corresponds to the projection-valued measure
$\big\{\ket{x}\bra{x};x\in\mathbb{Z}\big\}$ resulting in the
walker's state `collapsing' to position~$x$ on the lattice.
After~$t$ steps, the final state of the system
is~$\ket{\psi(t)}=U^t\ket{\psi(0)}$. The reduced state of the walker
is obtained by tracing out the coin.

The probability $P(x;t)$ that the walker will be found at the
position~$x$ is
\begin{equation}
    P(x;t)=\bra{x}\text{Tr}_\text{c}\left(\ket{\psi(t)}\bra{\psi(t)}\right)\ket{x},
\end{equation}
which is obtained by tracing out the coin of the walker-coin state
and then measuring the walker's position. We can characterize
$P(x;t)$ by the moments of this position distribution~$\langle
x^m\rangle$. The mean~$\langle x\rangle$ and variance $\langle
x^2\rangle-\langle x\rangle^2$ can be used as the measure of QWs and
show the signature of QWs compared to RWs. 
For a RW, the position variance $\sim t$, which is characteristic of
diffusive motion, whereas, for a QW, a quadratic enhancement is
achieved: the position variance~$\sim t^2$~\cite{Kem03}.

\subsection{Two walkers}
The extension to two independent quantum walkers is straightforward.
The new Hilbert space is~$\mathscr{H}\otimes\mathscr{H}$, and the
step operator is $U\otimes U$. The walkers evolve independently of
each other. To entangle the walkers and their respective coins
together, we change the step operator from $U\otimes U$ to
$\text{SWAP}^\tau(U\otimes U)$ for
\begin{equation}
    \text{SWAP}^\tau=\frac{1}{2}
        \begin{pmatrix}
            2 & 0 & 0 & 0\\
            0 & 1+(-1)^\tau& 1-(-1)^\tau & 0 \\
            0 & 1-(-1)^\tau & 1+(-1)^\tau & 0 \\
            0 & 0 & 0 & 2\\
        \end{pmatrix}
\end{equation}
expressed in the basis that naturally extends from the single-walker
case via the tensor product of Hilbert spaces.

The full initial state of the two-walker system is
$\ket{\psi(0)}\otimes\ket{\psi(0)}$ (each walker is initially
localized at the origin of their respective lines). After~$t$ steps,
the final state of the two-walker+coins system is
\begin{equation}
    \ket{\psi(t)}_{12}=U_2^t\ket{\psi(0)}_1\otimes\ket{\psi(0)}_2
\end{equation}
with~$U_2=\text{SWAP}^\tau(U\otimes U)$.
We now follow the evolution by performing a Fourier transform of the evolution operator to the ``momentum''~$k$ space.

The eigenvectors
\begin{equation}
    \ket{k}=\sum_x e^{ikx}\ket{x},
\end{equation}
of~$S$ and $S^\dagger$ in Eq.~(\ref{eq:S}) have the eigenrelations
\begin{equation}
    S\ket{k}=e^{-ik}\ket{k}, S^\dagger\ket{k}=e^{ik}\ket{k}
\end{equation}
for~$k$ a continuous real quantity.
The inverse transformation is
\begin{equation}
\ket{x}=\int^\pi_{-\pi}\frac{\text{d}k}{2\pi}e^{-ikx}\ket{k}.
\end{equation}
Each walker is initialized at the origin of the line so each walker's initial state is
\begin{equation}
    \ket{0}=\int^\pi_{-\pi}\frac{\text{d}k}{2\pi}\ket{k}.
\end{equation}
In the~$\{\ket{k}\}$ and~$\{\ket{j}\}$ bases for the two wakers, where~$j$ is a continuous index for the momentum of the second walker,
the evolution operator becomes
\begin{equation}
    U_2\ket{k}\otimes\ket{j}\otimes\ket{\Phi}_\text{c}=\ket{k}\otimes\ket{j}\otimes U_{kj}\ket{\Phi}_\text{c},
\end{equation}
with~$\ket{\Phi}_\text{c}$ the coin state and
\begin{equation}
U_{kj}=\text{SWAP}^\tau U_k\otimes U_j ,\end{equation} where
\begin{equation}
U_{k(j)}=\begin{pmatrix}
            e^{-ik(j)} & e^{-ik(j)}\\
            e^{ik(j)} & -e^{ik(j)} \\
        \end{pmatrix}.
\end{equation}
The~$k$ and~$j$ subscripts are of course the continuous walker-momentum eigenvalues.

The general density operator for the initial state of the system in
the $k$ basis can be expressed as
\begin{equation}
\rho(0)=\int_{-\pi}^\pi\frac{\text{d}k}{2\pi}\int_{-\pi}^\pi\frac{\text{d}k'}{2\pi}
\int_{-\pi}^\pi\frac{\text{d}j}{2\pi}\int_{-\pi}^\pi\frac{\text{d}j'}{2\pi}\ket{k}\bra{k'}\otimes\ket{j}\bra{j'}\otimes\ket{\Phi_0}\bra{\Phi_0}.
\end{equation} The final state after $t$ steps is
\begin{equation}
\rho(t)=\frac{1}{(2\pi)^4}\int_{-\pi}^\pi \text{d}k\int_{-\pi}^\pi
\text{d}k'\int_{-\pi}^\pi \text{d}j\int_{-\pi}^\pi
\text{d}j'\ket{k}\bra{k'}\otimes\ket{j}\bra{j'}\otimes
U_{kj}^t\ket{\Phi_0}\bra{\Phi_0}(U_{kj}^\dagger)^t.
\end{equation}
In terms of the superoperator $\mathcal{L}_{kk'jj'}\hat{O}=U_{kj}\hat{O}U_{kj}^\dagger$,
\begin{equation}
\rho(t)=\frac{1}{(2\pi)^4}\int_{-\pi}^\pi \text{d}k\int_{-\pi}^\pi
\text{d}k'\int_{-\pi}^\pi \text{d}j\int_{-\pi}^\pi
\text{d}j'\ket{k}\bra{k'}\otimes\ket{j}\bra{j'}\otimes\mathcal{L}^t_{kk'jj'}\ket{\Phi_0}\bra{\Phi_0}.
\end{equation}

The walkers' positions are on a two-dimensional integer lattice
labeled~$(x,y)$ with the initial position localized at~$(0,0)$.
Measurement of the first walker's position corresponds to the
projection-valued measure~$\{|x\rangle\langle
x|\otimes\one;x\in\mathbb{Z}\}$ on the two-walker reduced state
(after tracing out the coins), and the second walker's position
measurement is $\{\one\otimes|y\rangle\langle y|;y\in\mathbb{Z}\}$.
Joint measurement of the two walkers' positions corresponds to the
projection-valued measure
$$\{|x\rangle\langle x|\otimes|y\rangle\langle
y|;(x,y)\in\mathbb{Z}^2\}$$
acting on the reduced two-walker state.

The joint position distribution~$P(x,y;t)$ of finding the first
walker at~$x$ and the second at~$y$ is
\begin{equation}
    P(x,y;t)=\!_1\bra{x}_2\bra{y}\rho_\text{w}(t)\ket{y}_2\ket{x}_1,
\end{equation}
at time~$t$ where
\begin{equation}
    \rho_\text{w}(t)=\text{Tr}_\text{c}(\ket{\psi(t)}_{12}\bra{\psi(t)})
\end{equation} is the state of the two walkers obtained by tracing
out the coins. Thus the position distribution of each walker at
position~$x$ (the case for~$y$ is similar) is obtained by tracing
out the states of the two coins and the other walker
\begin{align}
\label{eq:distribution}
    P_1\left(x;t\right)=\bra{x}\text{Tr}_{2}\rho_\text{w}(t)\ket{x},\nonumber\\
P_2\left(y;t\right)=\bra{y}\text{Tr}_{1}\rho_\text{w}(t)\ket{y}.
\end{align}

The variance of the position distribution is especially important.
For standard QWs with single walkers, the variance of the position
distribution has been shown to evolve according to~$\sigma\propto
t^2$, whereas~$\sigma\propto t$ for RWs. This quadratic speedup of
spreading in a unitary evolution is a hallmark of the QW on a line.
For our case, namely QWs with two walkers switching their coin
partially for each step, the variance of each walker distribution
can still be used to qualify the quantum behavior of this kind walks.

If we trace out the second walker, the reduced density matrix is
\begin{align}
    \rho_1(t)&=\sum_y \text{Tr}\left\{\ket{y}\bra{y}\rho(t)\right\}\nonumber\\
    &=\frac{1}{(2\pi)^4}\int_{-\pi}^\pi \text{d}k\int_{-\pi}^\pi
    \text{d}k'\int_{-\pi}^\pi \text{d}j\int_{-\pi}^\pi
    \text{d}j'\ket{k}\bra{k'}\sum_y
    e^{-i(j-j')y}\text{Tr}\left(\mathcal{L}^t_{kk'jj'}\ket{\Phi_0}\bra{\Phi_0}\right).
\end{align}
The sum can be exactly carried out in terms of derivatives of the
$\delta$ function:
\begin{equation}
\frac{1}{2\pi}\sum_y y^m e^{-(j-j')y}=(-i)^m\delta^{(m)}(j-j').
\end{equation}
Inserting this result back into the expression of the reduced
density matrix yields
\begin{equation}
\rho_1(t)=\frac{1}{(2\pi)^3}\int_{-\pi}^\pi \text{d}k\int_{-\pi}^\pi
\text{d}k'\int_{-\pi}^\pi
\text{d}j\ket{k}\bra{k'}\text{Tr}\left(\mathcal{L}^t_{kk'j}\ket{\Phi_0}\bra{\Phi_0}\right).
\end{equation}
The probability for the first walker to reach a point $x$ at time
$t$ is
\begin{align}
P_1(x;t)&=\text{Tr}\left\{\ket{x}\bra{x}\rho_1(t)\right\}\nonumber\\
&=\frac{1}{(2\pi)^3}\int_{-\pi}^\pi \text{d}k\int_{-\pi}^\pi
\text{d}k'\int_{-\pi}^\pi \text{d}j
e^{-i(k-k')x}\text{Tr}\left(\mathcal{L}^t_{kk'j}\ket{\Phi_0}\bra{\Phi_0}\right).
\end{align}
Thus we can calculate the moments of this distribution:
\begin{align}
\langle x^m \rangle_t&=\sum_x x^m P(x;t)
\nonumber\\&\frac{(-1)^m}{(2\pi)^3}\int_{-\pi}^\pi
\text{d}k\int_{-\pi}^\pi \text{d}k'\int_{-\pi}^\pi \text{d}j
\delta^{(m)}(k-k')\text{Tr}\left(\mathcal{L}^t_{kk'j}\ket{\Phi_0}\bra{\Phi_0}\right).
\end{align}
For the first moment, we obtain
\begin{align}
\langle x\rangle_t&=\frac{-i}{(2\pi)^2}\int_{-\pi}^{\pi}
\text{d}k\int_{-\pi}^\pi \text{d}j
\frac{\text{d}}{\text{d}k}\text{Tr}\left(\mathcal{L}^t_{kj}\ket{\Phi_0}\bra{\Phi_0}\right)\nonumber\\
&=-\frac{1}{(2\pi)^2}\int_{-\pi}^{\pi}\text{d}k\int_{-\pi}^{\pi}\text{d}j\sum_{l=1}^t\text{Tr}\left(Z_1
\mathcal{L}^l_{kj}\ket{\Phi_0}\bra{\Phi_0}\right),
\end{align}
where $Z_1=\sigma_z\otimes\one$. We can carry out a similar
integration by parts to obtain the second moment:
\begin{equation}
\langle
x^2\rangle_t=\frac{1}{(2\pi)^2}\int_{-\pi}^{\pi}\text{d}k\int_{-\pi}^{\pi}\text{d}j\left\{\sum_{l=1}^t\sum_{l'=1}^l
\text{Tr}\left[Z_1\mathcal{L}^{l-l'}_{kj}\left(Z_1\mathcal{L}_{kj}^{l'}\ket{\Phi_0}\bra{\Phi_0}\right)\right]
+\sum_{l=1}^t\sum_{l'=1}^{l-1}
\text{Tr}\left[Z_1\mathcal{L}^{l-l'}_{kj}\left(\left(\mathcal{L}_{kj}^{l'}\ket{\Phi_0}\bra{\Phi_0}\right)Z_1\right)\right]\right\}.
\end{equation}

For $\tau=0$, the two-walker QW turns to be the standard Hadamard
walk for each walker. The first and second moments of the
distribution behave as same as those for the standard Hadamard walk.
That is, $\langle x\rangle_t$ is linear dependent on~$t$, while
$\langle x^2\rangle_t$ is quadratic dependent on~$t$.

For $0<\tau\leq1$, $U_{kj}$ is still unitary. For short time, the
moments of the position distribution behave similarly to those for
standard unitary walk. In the long-time limit we can solve the
moments of distribution analytically. Suppose $\mathcal{L}_{kj}$ is
a linear transformation and all the eigenvalues satisfy
$|\lambda|\leq1$. We drop the terms that will be zero at large~$t$
from the expressions for the moments.

As $\mathcal{L}_{kj}$ is linear, we can represent it as a
matrix acting on the space of $4\times 4$ operators. We choose the
representation to be
\begin{equation}
    \hat{O}=r_1\one^{12}+r_2\one^1\otimes\sigma^2_x+\cdot\cdot\cdot+r_{16}\sigma_z^1\otimes\sigma_z^2.
\end{equation}
The action of $\mathcal{L}_{kj}$ on~$\hat{O}$ is given
by the matrix
\begin{align}
\mathcal{L}_{kj}\hat{O}=M_{kj}
           (r_1 \; r_2 \cdots\;r_{16})
\end{align}
with none-zero matrix elements
\begin{align}
&(M_{kj})_{1,1}=(M_{kj})_{4,4}=(M_{kj})_{13,13}=(M_{kj})_{16,16}=
\frac{1}{4}\left[-(1+e^{i\pi \tau})\cos(k-j)+2\cos(k+j)\right],\nonumber\\
&(M_{kj})_{1,4}=(M_{kj})_{4,1}=(M_{kj})_{13,16}=(M_{kj})_{16,13}=
\frac{1}{8}e^{-i(k+j)}\left\{2+\left[1+(-1)^\tau\right]e^{2ij}-\left[1+(-1)^\tau+2e^{2ij}\right]e^{2ik}\right\},\nonumber\\
&(M_{kj})_{1,13}=(M_{kj})_{4,16}=(M_{kj})_{13,1}=(M_{kj})_{16,4}=
\frac{1}{8}e^{-i(k+j)}\left\{2-\left[1+(-1)^\tau\right]e^{2ij}+\left[1+(-1)^\tau-2e^{2ij}\right]e^{2ik}\right\},\nonumber\\
&(M_{kj})_{1,16}=(M_{kj})_{4,13}=(M_{kj})_{13,4}=(M_{kj})_{16,1}=
\frac{1}{4}\left[(1+e^{i\pi \tau})\cos(k-j)+2\cos(k+j)\right],\nonumber\\
&(M_{kj})_{6,6}=i(M_{kj})_{6,7}=-i(M_{kj})_{7,6}=(M_{kj})_{7,7}=i(M_{kj})_{10,6}=-(M_{kj})_{10,7}=(M_{kj})_{11,6}=i(M_{kj})_{11,7}\nonumber\\&=
-\frac{1}{4}\left[-1+(-1)^\tau\right]e^{i(j-k)},\nonumber\\
&(M_{kj})_{6,10}=i(M_{kj})_{6,11}=i(M_{kj})_{7,10}=-(M_{kj})_{7,11}=-i(M_{kj})_{10,10}=(M_{kj})_{10,11}=(M_{kj})_{11,10}=i(M_{kj})_{11,11}\nonumber\\&=
i\frac{1}{4}\left[-1+(-1)^\tau\right]e^{i(j-k)}.
\end{align}

In the integrand, the initial density matrix for the coins is
multiplied $l$ times by $\mathcal{L}_{kj}$ then multiplied on the left by
$Z_1$ and finally the trace is taken.
As $\text{Tr}\left\{\hat{O}\right\}=r_1+r_4+r_{13}+r_{16}$, $\langle
x\rangle_t$ is the same as multiplying the vector $
           (r_1 \; r_2 \; \cdots r_{16})^T$ $l$ times by $M_{kj}$ and then keeping only
the $r_1$, $r_4$,$r_{13}$ and $r_{16}$ components of the
result.
This gives the new expression of the first moment as
\begin{align}
\langle
x\rangle_t&=-\frac{1}{(2\pi)^2}\int_{-\pi}^{\pi}\text{d}k\int_{-\pi}^{\pi}\text{d}j
    \left(1 \; 0 \; 0 \; 1 \; 0 \;\cdots \; 0 \; 1 \; 0 \; 0 \; 1\right)
\sum_{l=1}^t
M_{kj}^l
    (r_1 \; r_2 \; \cdots \; r_{16})^T\nonumber\\
&=-\frac{1}{(2\pi)^2}\int_{-\pi}^{\pi}\text{d}k\int_{-\pi}^{\pi}\text{d}j
\sum_l^t \left(M_{kj}^l-2\ket{1}\bra{1}\otimes\one
M_{kj}^l\right)(r_1 \; r_2 \;\cdots\; r_{16})^T,
\end{align}
which is independent on~$t$ in the long time limit since
$M_{kj}^t\rightarrow 0$ with all the eigenvalues of $M_{kj}$
satisfying $0<|\lambda|\leq1$.
We then obtain
\begin{equation}
\langle x \rangle_t\approx
\left\{1-\frac{2}{(2\pi)^2}\int_{-\pi}^{\pi}\text{d}k\int_{-\pi}^{\pi}\text{d}j
\text{Tr}\left[\ket{1}\bra{1}\otimes\one(1-\mathcal{L}_{kj})^{-1}\mathcal{L}_{kj}\ket{\Phi_0}\bra{\Phi_0}\right]\right\}t+\text{oscillatory
terms}.
\end{equation}

For the second moment of the distribution, the superoperator
$\mathcal{L}_{kj}$ is unitary and preserves the identity,
$\mathcal{L}_{kj}\one=\one$. We will separate the traceless part
$\chi_0$ of the coins' initial state from the overall
state~\cite{BCA03,BCAprl},
\begin{equation}
r_1\one+(\ket{\Phi_0}\bra{\Phi_0}-r_1\one)\equiv r_1\one+\chi_0.
\end{equation}
then insert this result into the expression for
the second moment of the distribution to obtain
\begin{align}
\langle
x^2\rangle_t=&\frac{1}{(2\pi)^2}\int_{-\pi}^{\pi}\text{d}k\int_{-\pi}^{\pi}\text{d}j\Big\{\text{Tr}\left[\sum_{l=1}^t\mathcal{L}_{kj}^l(r_1\one+\chi_0)\right]
+2r_1\text{Tr}\left[Z_1\sum_{l=1}^t\sum_{l'=1}^{l-1}\mathcal{L}_{kj}^{(l-l')}Z_1\right]\nonumber\\
&+\sum_{l=1}^t\sum_{l'=1}^{l-1}\text{Tr}\left[Z_1\mathcal{L}^{l-l'}_{kj}\left(Z_1(\mathcal{L}^l_{kj}\chi_0)+(\mathcal{L}^l_{kj}\chi_0)Z_1\right)\right]\Big\}\approx
C_2t^2+\text{const.}+\text{oscillatory terms},
\end{align}
where \begin{equation}
C_2=1+\frac{1}{(2\pi)^2}\int_{-\pi}^{\pi}\text{d}k\int_{-\pi}^{\pi}\text{d}j\text{Tr}
\left[-2Z_1\left(1-\mathcal{L}_{kj}\right)^{-1}\mathcal{L}_{kj}\ket{\Phi_0}\bra{\Phi_0}\ket{1}\bra{1}\otimes\one
-4\ket{1}\bra{1}\otimes\one\left(1-\mathcal{L}_{kj}\right)^{-1}\mathcal{L}_{kj}\ket{\Phi_0}\bra{\Phi_0}\right].
\end{equation}

In brief the first and second moments of the position distribution
are linearly and quadratically dependent on~$t$ for $\tau=0$. For
$0<\tau\leq1$, in the long-time limit, a ballistic behavior for each
walker is obtained. Thus in the long-time limit the variance of the
position distribution for one of the walkers is quadratically
dependent on~$t$. In our case the walker cycles through a finite
sequence of coins (two) for each step, we still get ballistic
transport. If the coin is chosen randomly from step to step, i.e.
choosing coin from a infinite sequence for each step, a diffusive
spreading will be obtained~\cite{BCAprl,AVWW11,AAM+11}.

\section{Analysis}
In this section we analyze the two-walker state after reducing the
density matrix by tracing out the two coins. In
Subsec.~\ref{subsec:joint} we calculate the joint probability
distribution for two walkers to show the correlation introduced by
the SWAP operation on coins. In Subsec.~\ref{subsec:mutual} we
characterization classical and total correlation using the mutual
information (MI) and the quantum mutual information (QMI),
respectively, between the two coin-sharing walkers. In
Subsec.~\ref{subsec:corr}, we employ measurement-induced disturbance
(MID) as the measure of choice for quantifying quantum correlations
between the two walkers.

\subsection{Joint-position probability distribution for walkers}
\label{subsec:joint}

\begin{figure}
   \includegraphics[width=8.5cm]{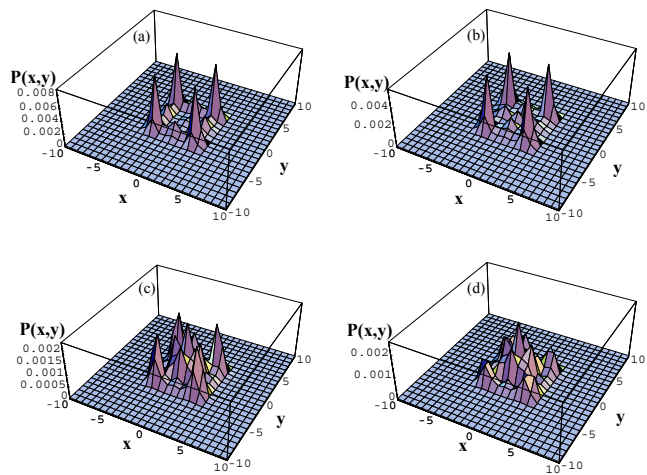}
   \caption{(Color online) The joint position distribution for a
   two-walker QW on a line after~$t=6$ steps with various
   SWAP-operator powers: (a) $\tau=0$, (b) $\tau=0.5$, (c) $\tau=0.8$,
   and (d) $\tau=1$.}
   \label{figure1}
\end{figure}

The joint probability distribution~$P(x,y;t)$ is calculated for
finding walker~$1$ in position~$x$ and walker~$2$ in position~$y$
for time~$t$ and shown in Fig.~\ref{figure1}.
For the power of the SWAP operation~$\tau=0$, which means the two
walkers are uncorrelated,
$P(x,y;t)$
is simply the product of
the two independent one-walker distributions,
$P_1(x;t)P_2(y;t)$,
where $P_1(x;t)$ is the probability distribution for finding walker
$1$ in position~$x$ after~$t$ and similarly for $P_2(y;t)$ and
walker~$2$.

The correlation functions $P(x,y;t=6)$ are shown in
Fig.~\ref{figure1}, The products of two independent single-walker
distributions is evident in Fig.~\ref{figure1}, which shows the
$\tau=0$ case. However, for $\tau>0$, the walkers are not
independent of each other due to the walkers (fractionally) swapping
their coins after each step.

Evidently the SWAP operation on the two coins correlates and perhaps
entangles the two coins. For classical RWs, the correlation between
the coins do not affect the walkers behavior and the position
distribution of each walker keeps the same as that of a standard RW
with one walker. Compared to RWs, the interference effect and
entanglement between each walker and his coin make the two-walker QW
behave differently. The correlation between the coins is transferred
to the walkers thereby inducing correlation between the two walkers.
However, the interference effect and entanglement between each
walker and his coin is partially destroyed by the SWAP operation.

In Fig.~\ref{figure1}, the joint position distributions of two
walkers after the $6^\text{th}$ step with various powers~$\tau$ of the SWAP
operation~$\tau$ displays key concepts. For $\tau=0$, the two walkers are
completely independent. For increasing~$\tau$, a small peak
appears in the middle of the distribution, which means there is
correlation between the two walkers. For $\tau=1$, the peak in the
middle gets large and the distribution still shows the quantum
behavior of two walkers. Thus the correlation between two walkers
increases with increasing power~$\tau$ of the SWAP operation.


\subsection{Mutual information and quantum mutual information}
\label{subsec:mutual}

In probability theory and information theory, the mutual information of two random
variables quantifies mutual dependence of the two random variables.
Here we use the mutual information as a measure of classical correlation between two
quantum walkers fractionally swapping their coins.
Formally the mutual information of the positions of the two walkers can be defined
as
\begin{align}
    I_\text{c}(t)=\int\int P(x,y;t)\log_2\left[\frac{P(x,y;t)}{P_1(x;t)P_2(y;t)}\right]\text{d}x\text{d}y.
\label{eq:MI}
\end{align}
As there is no classical correlation, the mutual information of two independent
walkers is zero.

\begin{figure}
   \includegraphics[width=8.5cm]{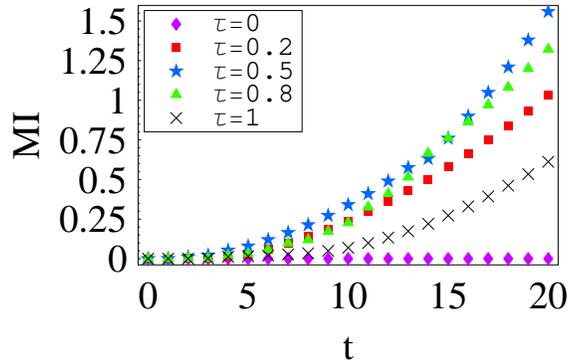}
   \caption{(Color online) The mutual information (MI) for
   the positions of the two walkers as a function of time~$t$ with various
   choices of the SWAP operator power~$\tau$.}
   \label{fig:MI}
\end{figure}

The SWAP operation between two coins connects the two walkers and
introduces correlations. We use the mutual information as a measure of classical
correlations between two walkers. In Fig.~\ref{fig:MI} the mutual information of the
two walkers' positions is depicted for various powers of the SWAP
operation~$\tau$.
In the case of $\tau=0$, the two walkers, each holding an independent Hadamard
coin, walk on a separate line independently of the other's dynamics so the mutual information of the
two walkers' positions is necessarily zero.

For a classical RW, if two walkers exchange their coin after every
step, any correlation between each walker and his coin is severed.
However, for the QW, correlations between the two walkers are not
completely destroyed. For $\tau=1/2$ the two coins are maximally
entangled by the $\sqrt{\text{SWAP}}$ operation, thereby maximizing
the correlation between the two walkers.

In addition to classical correlations, the SWAP operation on two
coins introduces quantum correlations between the two walkers as
well. Now we consider the total correlation between two quantum
walkers on a line under the influence of the SWAP operation between
the coins. We use the quantum mutual information as a measure of the
total correlation including classical and quantum
correlations~\cite{L08,LL07,L03,L06,OZ01,V03,M10,GBB10} between the
walkers.

Given a bipartite state~$\rho_\text{w}$ and the reduced density
matrices denoted by~$\rho_1$ and~$\rho_2$, the quantum mutual information,
\begin{equation}
    I\left[\rho_\text{w}(t)\right]
        =S\left[\rho_1(t)\right]+S\left[\rho_2(t)\right]-S\left[\rho_\text{w}(t)\right],
\label{eq:mutual}
\end{equation}
is a reasonable measure for total correlation between systems~$1$ and~$2$,
where~$S(\bullet)$ denotes von Neumann entropy. The quantum mutual information can be
interpreted analogously to the classical case, namely
\begin{equation}
    I\left[\rho_\text{w}(t)\right]=S\left[\rho_\text{w}(t)\|\rho_{1}(t)\otimes\rho_{2}(t)\right],
\label{eq:QMI}
\end{equation}
where $S(\bullet\|\bullet)$ denotes quantum relative entropy.
Intuitively, the quantum mutual information reports the shared quantum information between walkers
$1$ and~$2$.

Quantum mutual information is a good measure for quantifying the reduction of the uncertainty of one
variable through knowing the other variable.
For example, if walkers~$1$
and~$2$ are independent, then knowing walker~$1$'s position does not
give any information about $2$'s position and vice versa so their
quantum mutual information is zero. Otherwise, if walkers~$1$ and~$2$ are correlated, the
quantum mutual information is positive, and the bound on this correlation is approximately $2\log_2d$ with~$d$
the dimension of the walker state.

\begin{figure}
   \includegraphics[width=8.5cm]{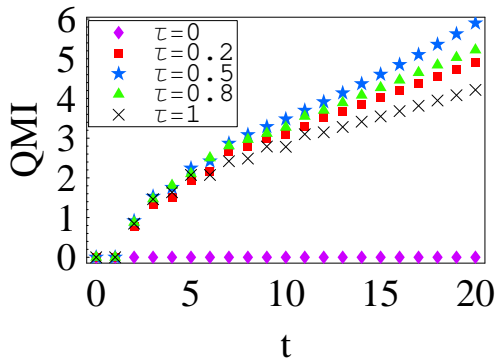}
   \caption{(Color online) The quantum mutual information (QMI) for
   two walkers' positions as a function of time~$t$ with various
   powers
   of the SWAP operator~$\tau\in \left[0,1\right]$.}
   \label{fig:QMI}
\end{figure}

The total correlation between two walkers increases with the powers~$\tau$
of the SWAP operations as does the classical correlation.
Figure~\ref{fig:QMI} presents plots of the quantum mutual information for an evolving
two-walker QW system with various powers of the SWAP operation.
For fixed value of~$\tau$, the quantum mutual information increases monotonically as the
number of the steps increases. For $\tau=0$, the quantum mutual informations stays at zero,
which means the total correlations are cut off between the two
walkers. The quantum mutual informations for $\tau>0$ are positive and a maximum for
$\tau=1/2$. We can see the maximal correlation (both total
correlation and classical correlation) between walkers where we
expect to see the coins entangled maximally.

\subsection{Quantum correlations}
\label{subsec:corr}

Quantum discord~\cite{OZ01} is a popular measure for characterizing
purely quantum correlations within a bipartite state, but evaluating
quantum discord requires extremizing over local measurement
strategies. Such an evaluation can be intractable so
measurement-induced disturbance was proposed as an operational
measure that avoids such an onerous optimization.
measurement-induced disturbance relies on diagonalizing the reduced
density operators, which is tractable for small systems. For
standard QWs, measurement-induced disturbance acts as a loose upper
bound on quantum discord and tends to reflect well behavioral trends
of the quantum discord~\cite{SBC10}.

Measurement-induced disturbance and quantum discord have been
compared for noisy linear and cyclic QWs~\cite{RSCB11}. Whereas
measurement-induced disturbance has an operational definition,
operational meaning for quantum discord is less
straightforward~\cite{CAB+11}. On the other hand measurement-induced
disturbance tends to overestimate non-classicality because it does
not optimize over local measurements. Despite this over-estimation,
measurement-induced disturbance gives a loose upper bound on quantum
discord and reflects well quantum correlations for
QWs~\cite{RSCB11}. Nevertheless, care is need in interpretations of
quantum correlations from measurement-induced disturbance as regimes
exist where lack of optimization leads to overestimates of
non-classicality.

Given the reduced two-walker bipartite state $\rho_\text{w}$
(tracing over coins), let the reduced density matrices for the two
walkers be diagonalized to
\begin{equation}
    \rho_i=\sum_j p_i^j\Pi_i^j,
\end{equation}
for $i=1,2$, where $\{\Pi_i^j\}$ is a complete projection-valued
measure (i.e.\ using von Neumann measurements) for each~$i$. Summing
over joint projections on the two-walker state yields the
diagonalized state
\begin{equation}
    \Pi\rho_\text{w}(t)=\sum_{j,k}\Pi_1^j\otimes\Pi_2^k\rho_\text{w}(t)\Pi_1^j\otimes\Pi_2^k.
\end{equation}
The leftmost $\Pi$ is an operator on the density matrix that
diagonalizes it in the spectral basis corresponding to $\Pi_i^j$
projective measurements. The operator~$\Pi$ can also be described as
a `local measurement strategy'.

Correlations between the two reduced-walker states $\rho_1$
and~$\rho_2$ are regarded as classical if there is a unique local
measurement strategy $\Pi$ leaving $\Pi\rho_\text{w}(t)$ unaltered
from the original two-walker state~$\rho_\text{w}$~\cite{L08}. We
ascertain whether the reduced two-walker state is `quantum' by
determining whether a local measurement strategy exists that leaves
the two-walker state unchanged.

The degree of quantumness is given by the measurement-induced disturbance~\cite{L08}
\begin{equation}
\label{eq:MID}
    Q\left[\rho_\text{w}(t)\right]=I\left[\rho_\text{w}(t)\right]-I\left[\Pi\rho_\text{w}(t)\right],
\end{equation}
for $I(\bullet)$ the quantum mutual information. By
construction~$\Pi\rho_\text{w}(t)$ is classical. Hence
$I\left[\Pi\rho_\text{w}(t)\right]$ quantifies the classical
correlations in $\rho_\text{w}(t)$, which must
equal~$I_c$~(\ref{eq:MI}). Thus, the measurement-induced disturbance
is the difference between the quantum and classical mutual
information, which quantify total and classical correlations.
Accordingly, Eq.~(\ref{eq:MID}) is interpreted as the difference
between the total and classical correlations, which are represented
by the quantum mutual information and the mutual information.
\begin{figure}
   \includegraphics[width=8.5cm]{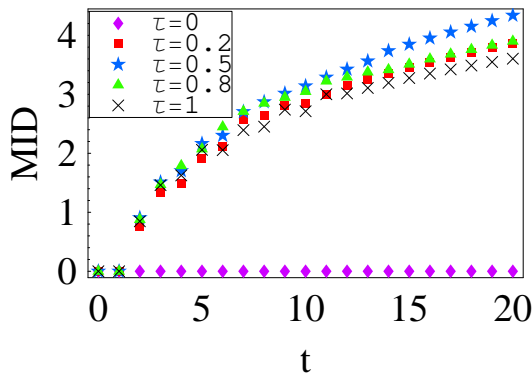}
   \caption{(Color online) The measurement-induced disturbance (MID) for
   two walkers' positions as a function of $t$ with
   various powers of the SWAP operation~$\tau\in \left[0,1\right]$.}
   \label{fig:MID}
\end{figure}

Figure~\ref{fig:MID} presents plots of measurement-induced
disturbance of two walkers in a two-walker QW system. From the
relationship between total, classical and quantum correlations
discussed in the previous paragraph, we expect a tight quantitative
relationship between Figs.~\ref{fig:MI}, \ref{fig:QMI} and
\ref{fig:MID}. We see that the total correlation represented by
quantum mutual information in Fig.~\ref{fig:QMI} dwarfs the
classical correlation represented by mutual information in
Fig.~\ref{fig:MI}. Hence Figs.~\ref{fig:QMI} and~\ref{fig:MID} are
quite similar.

Pure quantumness, represented by the measurement-induced disturbance
depicted in Fig.~\ref{fig:MID}, is  monotonically increasing with
$\tau$, and pure quantum correlations thus exist between the two
walkers. This pure quantum correlation is not due to direct
walker-walker interactions but are rather due to the walkers
fractionally swapping their coins after each step, and the pure
quantum correlations survive the tracing out of these coins.

\section{Conclusions}

We analyze the dynamics and entanglement of two quantum walkers who
fractionally swap (i.e.\ perform a SWAP$^\tau$ unitary operation
with $0\leq\tau\leq 1$) coins. We use mutual information, quantum
mutual information and measurement-induced disturbance as measures
to isolate classical vs quantum correlations.

Quantum discord would be a valuable measure to use but is not
tractable in our case. However, measurement-induced disturbance
suffices to show that pure quantum correlations are induced by
having the walkers fractionally swap coins and then tracing out the
degrees of freedom for the two coins. In fact both classical and
quantum correlations co-exist between the two walkers for $\tau>0$.
Quantum correlations are strongest for $\tau=1/2$ where the
$\sqrt{\text{SWAP}}$ has maximal entangling power and is applied
after each step.

Fractional swapping is a quintessentially quantum process.
Classically the walkers can swap their coins or not, or they could
SWAP their coins sometimes, either deterministically or
probabilistically. In our two-walker fractional-coin-swapping
scenario, a fractional coin swap is effected after every step
identically. Classically one would expect that coin swapping would
not affect the dynamics anyway because the coin state is
unimportant: the flip chooses a random outcome. Our classical
understanding of coin swapping thus gives little intuition about the
quantum case. Thus, our analysis is quite valuable in that we
characterize this two-walker fractional-coin-swapping case carefully
and devise appropriate, tractable, meaningful classical and quantum
correlation measures to study entanglement for this system.

\acknowledgments
We thank C. Di Franco, A. Ahlbrecht, A. H.
Werner and R. F. Werner for critical comments. PX acknowledges
financial support from the National Natural Science Foundation of
China under Grant Nos 11004029 and 11174052, the Natural Science
Foundation of Jiangsu Province under Grant No BK2010422, the Ph.D.\
Program of the Ministry of Education of China, the Excellent Young
Teachers Program of Southeast University and the National Basic
Research Development Program of China (973 Program) under Grant No
2011CB921203. BCS acknowledges financial support from CIFAR, NSERC
and AITF.

\end{document}